\documentclass[conference]{IEEEtran}
\IEEEoverridecommandlockouts

\usepackage{tikz}
\usepackage{cite}
\usepackage{amsmath,amssymb,amsfonts}
\usepackage{algorithmic}
\usepackage{graphicx}
\usepackage{textcomp}
\def\BibTeX{{\rm B\kern-.05em{\sc i\kern-.025em b}\kern-.08em
    T\kern-.1667em\lower.7ex\hbox{E}\kern-.125emX}}
    
\newcommand\copyrighttext{%
\footnotesize Copyright and Reprint Permission: Abstracting is permitted with credit to the source. Libraries are permitted to photocopy beyond
the limit of U.S. copyright law for private use of patrons those articles in this volume that carry a code at the bottom of the first page,
provided the per-copy fee indicated in the code is paid through Copyright Clearance Center, 222 Rosewood Drive, Danvers, MA 01923. For reprint or republication permission, email to IEEE Copyrights Manager at pubs-permissions@ieee.org. All rights
reserved. Copyright ©2019 by IEEE.}
\newcommand\copyrightnotice{%
\begin{tikzpicture}[remember picture,overlay]
\node[anchor=south,yshift=10pt] at (current page.south) {\fbox{\parbox{\dimexpr\textwidth-\fboxsep-\fboxrule\relax}{\copyrighttext}}};
\end{tikzpicture}%
}
\begin{document}
\title{A Comparison of Semantic Similarity Methods for Maximum Human Interpretability\\
}

\author{\IEEEauthorblockN{Pinky Sitikhu\textsuperscript{1}, Kritish Pahi\textsuperscript{2},  Pujan Thapa\textsuperscript{3}, Subarna Shakya\textsuperscript{4}}
\IEEEauthorblockA{Department of Electronics and Computer Engineering, Tribhuwan University\\
Pulchowk Campus, IOE, Nepal\\
\textsuperscript{1}pinky.sitikhu524@gmail.com\\
\textsuperscript{2}kritishpahi@gmail.com\\
\textsuperscript{3}pujanthapa88.63@gmail.com\\
\textsuperscript{4}drss@ioe.edu.np}
}

\maketitle

\begin{abstract}
The inclusion of semantic information in any similarity measures improves the efficiency of the similarity measure and provides human interpretable results for further analysis. The similarity calculation method that focuses on features related to the text's words only, will give less accurate results. This paper presents three different methods that not only focus on the text's words but also incorporates semantic information of texts in their feature vector and computes semantic similarities. These methods are based on corpus-based and knowledge-based methods, which are: cosine similarity using tf-idf vectors, cosine similarity using word embedding and soft cosine similarity using word embedding. Among these three, cosine similarity using tf-idf vectors performed best in finding similarities between short news texts. The similar texts given by the method are easy to interpret and can be used directly in other information retrieval applications.

\end{abstract}


\section{Introduction}
Text similarity has been one of the most important applications of Natural Language Processing. Two texts are said to be similar if they infer the same meaning and have similar words or have surface closeness. Semantic similarity measures the degree of semantic equivalence between two linguistic items, be they concepts, sentences, or documents\cite{b15}. A proper method that computes semantic similarities between documents will have a great impact upon different NLP applications like document classification, document clustering, information retrieval, machine translation, and automatic text summarization. There are different approaches to compute similarities between documents that use lexical matching, linguistic analysis, or semantic features. In the context of short texts, methods for lexical matching might work for trivial cases\cite{b3}. But it is arguably not an efficient method because this method considers whether the words of short texts look-alike eg. in terms of distances, lexical overlap\cite{b1} or largest common substring\cite{b2}. Generally, linguistic tools like parsers and syntactic trees are used for short text similarity, however, all the texts such as tweets might not be parsable. A linguistically valid interpretation might be very far from the intended meaning. So, semantic features are quite important in text mining. For semantic features, external sources of structured semantic knowledge such as Wikipedia, WordNet, or word embeddings are used\cite{b3}.
\par
The main objective of this paper is to compare the semantic similarities between short texts to maximize human interpretability. The basic idea to compute text similarities is by determining feature vectors of the documents and then calculating the distance between these features. A small distance between these features means a high degree of similarity, whereas a large distance means a low degree of similarity. Euclidean distance, Cosine distance, Jensen Shannon Distance, Word Mover distance\cite{b4} are some of the distance metrics used in the computation of text similarity. This paper presents two different methods to generate features from the documents or corpus: (1) using Tf-idf vectors and (2) using Word Embeddings and presents three different methods to compute semantic similarities between short texts: (1) Cosine similarity with tf-idf vectors (2) Cosine similarity with word2vec vectors (3) Soft cosine similarity with word2vec vectors.
\copyrightnotice
\section{Related Work}
Computation of similarity between the texts has been an important method of data analysis which can be further used in different NLP applications like information retrieval, sentiment analysis. So, there has been wide research for this computation using different features of texts and implementing different algorithms and metrics. Different case studies have also been carried out showing various applications of semantic similarity.

N. Shibata et. al \cite{b14} performed a comparative study to measure the semantic similarity between academic papers and patents, using three methods like the Jaccard coefficient, cosine similarity of tf-idf vector, and cosine similarity of log-tf-idf vectors. All of these methods are corpus-based methods and they also performed a case-study for further analysis. S. Zhang et. al\cite{b15} explored the different applications of semantic similarity related to the field of Social Network Analysis. 

H. Pu et. al\cite{b16} incorporated a semantic similarity calculation algorithm that used the large corpus to compare and analyze several words. This method used the Word2Vec model to calculate semantic information of the corpus and used Li similarity measure to compute similarity. Wael H. Gomaa\cite{b18} also discussed three text similarity approaches: string-based, corpus-based and knowledge-based similarities and proposed different algorithms based on it.

The above works showed that different methods and metrics have already been carried out for this research. In this experiment, we have also introduced three different semantic similarity measures based on both corpus-based and knowledge-based methods. The basic concept was that the small features from text words are not enough, and the inclusion of word semantic information in the process can improve the method. The tf-idf vectors were used to gain information about the corpus or document in case of a corpus-based method ie. cosine similarity using tf-idf vectors. The knowledge-based method included computation of word embedding for each text, and cosine and soft-cosine similarity metrics were used for similarity calculation between these word embedding vectors.

\section{Methodology}

\subsection{Dataset}

The dataset used in this experiment is the popular and publicly available short news articles called AG`s news topic classification.  
\subsubsection{AG`s News Topic Classification Dataset}
This news corpus consists of news articles from the AG`s corpus of news articles\cite{b5} on the web and have 4 largest classes. Each class consists of 30,000 training samples and 1900 testing samples. This dataset is provided by the academic community for research purposes in data mining, information retrieval, etc.

\subsection{Data Cleaning and Preprocessing}

Basic preprocessing steps like lemmatization, stop-word removal, and filtering characters, punctuations and numbers were done. WordNetLemmatizer was used to lemmatize the words and the English stop-words were taken from NLTK library\cite{b7}.

\subsection{Feature Vectors}
In machine learning, a feature vector is a n-dimensional vector of numerical features that represent some object or context. It holds an important place in computing semantic similarities between texts. In this experiment, two different methods have been used to compute feature vectors. They are:

\subsubsection{Tf-idf Vectors}
TF-IDF is an abbreviation for Term Frequency-Inverse Document Frequency and is a very  common algorithm to transform a text into a meaningful representation of numbers. Tf-idf weight is a statistical measure that evaluates the importance of a particular word to a document.

Mathematically,
\begin{equation}
    tfidfweight = \sum_{i\in_d} tf_{i,d} * log(\frac{N}{df_i})
\end{equation}

where,
\(tf_{i,d}\) is the number of occurrences of \(i^{th}\) term in document d, \(df_i\) is the number of documents containing \(i^{th}\) term, N is the total number of documents.
\par
A tf-idf model was created using sklearn vectorizer model. This model was fitted using the documents and a set of tf-idf vectors containing tf-idf weight of each words of the documents were created. Now, these tf-idf vectors were used as a feature vectors for measuring similarities between the news dataset.

\subsection{Word Embeddings}
Word embeddings are vector representations of a word obtained by training a neural network on a large corpus. It has been widely used in text classification using semantic similarity. Word2vec is one of the most widely used forms of word embeddings. The word2vec takes text corpus as input and produces word vectors as output, which can be further used to train any other word to obtain its corresponding vector value. This word2vec model uses a continuous skip-gram model\cite{b8}, based on the distributional hypothesis.
An open-source library, Gensim\cite{b9} provides different pre-trained word2vec models trained on different kinds of datasets like GloVe, google-news, ConceptNet, Wikipedia, twitter\cite{b9}. A pre-trained word embeddings named ConceptNet Numberbatch 19.08\cite{b11} was used as a word2vec model to create feature vectors for the dataset in this experiment.

\subsection{Similarity Measures}
Similarity function or measure is a real-valued function that quantifies the similarity between two objects. This experiment presents two similarity measures: cosine similarity and soft-cosine similarity.

\subsubsection{Cosine Similarity}
It is a similarity measure which measures the cosine of the angle between two vectors projected in a multi-dimensional plane. It is the judgment based on orientation rather than magnitude. 

Given two vectors of attributes, A and B, the cosine similarity is presented using a dot product and magnitude as:

\begin{equation}
    similarity = cos(\theta) = \frac{A.B}{|A| |B|}
\end{equation}

\subsubsection{Soft Cosine Similarity}
A soft cosine or soft similarity between two vectors generalizes the concept of cosine similarity and considers similarities between pairs of features\cite{b12}. The traditional cosine similarity considers the Vector Space Model (VSM) features as independent or completely different, while soft cosine measure considers the similarity of features in VSM as well. The similarity between a pair of features can be calculated by computing Levenshtein distance, WordNet similarity or other similarity measures. For example, words like ``cook`` and ``food`` are different words and are mapped to different points in VSM. But, semantically they are related to each other. 

Given two N-dimensional vectors a and b, the soft cosine similarity can be calculated as follows:

\begin{equation}
    softcosine(a, b) = \frac{\sum_{i,j}^N s_{ij} a_i b_j}{\sqrt{\sum_{i, j}^N s_{ij} a_i a_j} \sqrt{\sum_{i, j}^N s_{ij} b_i b_j}}
\end{equation}

where \(s_{ij}\) = similarity(feature, feature)
\vskip 0.05in
The matrix s represents the similarity between features. If there is no similarity between features (\(s_{ii}\) = 1, \(s_{ij}\) = 0 for i \(\neq\) j) then, this equation is equivalent to cosine similarity.

\section{Proposed Method}
On the basis of methods to compute feature vector and similarity measures, this paper presents three methods to compute similarities between short text data.

\subsection{Cosine Similarity using TF-IDF Vectors}
The pre-processed documents were converted into tf-idf vectors by using a vectorized tf-idf model. The obtained vectors were a sparse matrix containing tf-idf weights for each word of each document having the size of [number of documents * number of features(unique words)]. Now, these tf-idf weights from the matrix were used as a feature for each document, and similarity between documents are computed using cosine similarity. The inbuilt cosine similarity module from sklearn was used to compute the similarity.

\subsection{Cosine Similarity using Word2Vec Vectors}
In this method, the pre-trained word2vec model was loaded using gensim\cite{b9}. This word2vec model was used to compute the vector values or word embeddings of each word of all the preprocessed documents. The word2vec vectors for the words that do not exist in the vocabulary of the pre-trained model were set to zero. An average of all the word vectors of a document was computed and the resultant vector was used as a feature word2vec vector for that document. So, all the documents were vectorized from n-gram vectors into word2vec vectors. These word2vec vectors were then used as feature vectors to compute the cosine similarity within the documents. 

\subsection{Soft Cosine Similarity using Word2Vec Vectors}
This method also used the word embeddings from same pre-trained word2vec model. These word2vec vectors were used to construct a term similarity index which computes cosine similarities between these word2vec vectors from the pre-trained model.

The pre-processed documents were converted into a dictionary or a bag of words model and the formed dictionary was converted into a corpus (a sparse vector containing unique word ids and its number of occurrences). Since soft cosine measure uses a similarity matrix between pair of features, the dictionary (feature of the dataset) and term similarity index (features of word2vec model) were used to compute a term similarity matrix. Finally, soft cosine similarity between the documents was computed. The modules provided by gensim\cite{b9} were used to compute this similarity measure.

\section{Result and Analysis}
About 2000 news articles from the dataset were selected randomly for the experiment. The similarity of each news article was computed against itself and all the other articles. A huge matrix of size (number of news articles * number of news articles) ie. 2000 * 2000 was created. We know that self-similarity is always 1. So, all the diagonal values of the similarity matrix were replaced by 0 because we wanted to find the most similar document for any document beside itself. The most similar news article for a news article was computed by finding out the maximum similarity value in the similarity matrix.
\par
These figures show the illustration of the procedure explained above. D1, D2, D3, D4, D5  represents five documents. 
\begin{figure}[htbp]
\centerline{\includegraphics[scale=0.6]{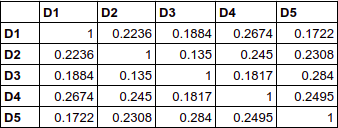}}
\caption{Similarity Matrix of five documents with each other.}
\label{fig}
\end{figure}

\begin{figure}[htbp]
\centerline{\includegraphics[scale=0.6]{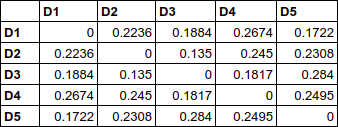}}
\caption{Replacing all diagonal values (self-similarity values) to 0.}
\label{fig}
\end{figure}

\begin{figure}[htbp]
\centerline{\includegraphics[scale=0.6]{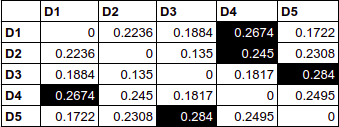}}
\caption{Finding out the most similar document on the basis of similarity value.}
\label{fig}
\end{figure}

The analysis of the performance of these three methods was done by retrospective analysis by measuring the top-1 accuracy. Top-1 accuracy is the conventional accuracy in which the answer from the model must be the expected answer. So, the newsgroup or news category of each news article and its respective most similar news articles were checked and compared. If an article and its most similar article are in the same category or newsgroup, then it is noted as correctly classified, otherwise not. The accuracy of each method was calculated on the basis of this rule, ie.

\begin{equation}
    accuracy = \small{\frac{\text{total no. of correctly classified news articles}}{\text{total number of news articles}} * 100}
\end{equation}

The table below shows the top-1 accuracy of each method using the dataset.
\begin{table}[htbp]
\caption{Table Showing accuracy of each method}
\begin{center}
\scalebox{0.95}{
\begin{tabular}{|c|c|}
\hline

\textbf{Methods used to calculate text similarity} & \textbf{\textit{Top-1 Accuracy (in $\%$)}} \\
\hline
Cosine Similarity using tf-idf Vectors & 76.8 \\
\hline
Cosine Similarity using Word2Vec Vectors & 75.9 \\
\hline
Soft Cosine Similarity using Word2Vec Vectors & 76.05\\
\hline
\end{tabular}}
\label{tab1}
\end{center}
\end{table}

This table shows that all of the three methods performed quite well with great accuracy. But, among these three methods, cosine similarity using tf-idf showed greater accuracy. Cosine similarity with word2vec had relatively low accuracy among all three methods. The reason behind this is the fact that the document vector is computed as an average of all word vectors in the document and the assignment of zero value for the words, that are not available in word2vec vocabulary. Soft cosine similarity performed better than cosine similarity with word2vec as this method used a similarity matrix with pre-trained word embeddings and dictionary or BoW of the documents as features. 

The obtained result has confirmed that the inclusion of semantic information in the form of tf-idf vectors and word embeddings, provides positive improvement in the method of computing semantic similarity of short texts. 

\section{Conclusion}
In this research paper, we performed a comparison between three different approaches for measuring the semantic similarity between two short text news articles. AG`s news-related data sets were used to verify the experiment. The three approaches are Cosine similarity with tf-idf vectors, cosine similarity with word2vec vectors, soft cosine similarity with word2vec vectors.  All of these three methods had shown promising results. Among these three vectors, cosine with tf-idf had the highest accuracy when the results were cross-validated and the newsgroup of a news article and its corresponding most similar article were compared. The most similar documents given by the method are easy to interpret, which makes it easier to use in different kinds of information retrieval methods. 

The accuracy of the other two methods can be increased by using the Doc2Vec model instead of the Word2Vec model, which represents a document as a vector and does not average the word vectors of a document. This model can be implemented for further improvement of the methods.

\section*{Acknowledgment}
We would like to express our deepest gratitude to Prof. Dr. Subarna Shakya for supervising and motivating us for this research work. We would like to thank Mr. Manish Munikar and Mr. Nishan Pantha for providing us with valuable feedback for this work. Also, we are immensely grateful to all the people who have directly or indirectly supported us in this work.

\end{document}